\documentclass[aps,pra,showpacs,twocolumn,groupedaddress,floatfix,longbibliography]{revtex4-1}

\usepackage{enumerate}
\usepackage{graphicx}
\usepackage{amssymb}
\usepackage{pifont}
\usepackage{amsmath}
\usepackage{multirow}
\usepackage{minibox}
\usepackage{enumerate}
\usepackage{mathtools}
\usepackage{wasysym}

\begin{document}

\def\crta{\vrule height1.41ex depth-1.27ex width0.34em}
\def\dj{d\kern-0.36em\crta}
\def\Crta{\vrule height1ex depth-0.86ex width0.4em}
\def\Dj{D\kern-0.73em\Crta\kern0.33em}
\dimen0=\hsize \dimen1=\hsize \advance\dimen1 by 40pt

\title{Response to ``Retraction Note: Can Two-Way Direct Communication
  Protocols Be Considered Secure?''}

\author{Mladen Pavi{\v c}i{\'c}}

\email{mpavicic@irb.hr}

\affiliation{Department of Physics---Nanooptics, 
Faculty of Math. and Natural Sci.~I,
Humboldt University of Berlin, Germany and\\
Center of Excellence for Advanced Materials
and Sensing Devices (CEMS), Photonics and Quantum Optics Unit,
Ru{\dj}er Bo\v skovi\'c Institute, Zagreb, Croatia}

\begin{abstract}
  This is a response to ``Retraction Note: Can Two-Way Direct
  Communication Protocols Be Considered Secure? Mladen
  Pavi{\v c}i{\'c},'' Nanoscale Research Letters, {\bf 14}, 242
  (2019). The ``anonymised'' report which served the editors to
  retract the paper is reproduced in full. A response to the
  report by the author of the original paper, sent to the editors
  of the journal, is also reproduced in full. The author analyses
  and discusses both the report and the retraction note and
  explains why he cannot agree to the retraction. 
\end{abstract}



\maketitle

\section{Introduction}
\label{intro}

The editors of the {\em Nanoscale Research Letters} recently published
the following Retraction Note:

\medskip
\parindent=0pt
``Retraction Note: Can Two-Way Direct Communication Protocols
Be Considered Secure?''

{\em Nanoscale Research Letters}, {\bf14}, 242 (2019) 

{\tt https://doi.org/10.1186/s11671-019-3086-8}

\medskip

The editors have retracted this article [1] because after publication 
concerns were raised regarding the validity of the conclusions drawn. 
Post-publication peer review has revealed a flaw in the application 
of the key rate equation $r=I_{AB}-I_{AE}$. For calculation of the
term $I_{AE}$, the effect of disturbance ($D$) on both the message
mode (MM) and control mode (CM) was not taken into account. The main
claim of the paper cannot be reliably reached. The author does not
agree to this retraction.

\smallskip
1. Pavi{\v c}i{\'c}, M. (2017), Can Two-Way Direct Communication Protocols 
Be Considered Secure? {\em Nanoscale Res. Lett.} {\bf 12}:552. 

\parindent=20pt

\bigskip

Reportedly, the editors decided to obtain a post-publication peer
review report of the original article that then served them to
retract the article. The report is given in Sec.~\ref{report}.

\section{ Anonymised Report}
\label{report}

The original article deals with the ping-pong protocol. These type 
of protocols were originally sold as a way to distribute secret 
keys directly (without the need for two-way communication involving
error correction and privacy amplification), but it is by now widely
recognized in the community that this was not a useful direction. 
Hence, the author of the article, and also the authors of the 
comment on that article, treat the protocol as a specific form of a 
QKD protocol, which performs error correction and privacy 
amplification after the initial exchange of quantum signals and 
subsequent measurements.

I fully agree with the comments made by the comment authors
\cite{shaari-mancini-pirandola-lucamarini-18-v1,shaari-mancini-pirandola-lucamarini-18-v2,shaari-mancini-pirandola-lucamarini-18-v3}. 
The author of the original article \cite{pavicic-nrl-17}
misunderstood the origin of the key rate formula $r=I_{AB}-I_{AE}$.
(Note that the security definitions changed since the late 1990s,
and we would not write this type of formula anymore.) But for the
sake of argument, let's stick to the old notions: The information
is always the information on the raw data. And while $I_{AB}$
directly comes from the observed error rate 
of the raw data (disturbance D of the message mode (MM)) in this 
case, the term $I_{AE}$ needs to be calculate from all available 
constraints, including the disturbance in the MM and also the control
mode CM. 

For the given Man-in-the-middle attack, we would create no 
disturbance in the MM mode, but full disturbance (random outcome) in 
the CM mode. Therefore, one would conclude that $I_{AB}=1$ (no 
disturbance in the MM mode) and also $I_{AE}=1$, since the disturbance 
of CM shows that Eve could have performed the attack, thus obtain all
signal, and thus the key rate would be zero for this attack. This is 
contained in the security analyses of these protocols. For smaller 
disturbance rate, the key rate turns positive, utilizing the privacy 
amplification method. Again, we are doing QKD here, where privacy 
amplification is allowed... 

I will not go in more detail on the discussion between the author
\cite{pavicic-nrl-17-response} of the article \cite{pavicic-nrl-17}
and the authors of the comment
\cite{shaari-mancini-pirandola-lucamarini-18-v1,shaari-mancini-pirandola-lucamarini-18-v2,shaari-mancini-pirandola-lucamarini-18-v3}
regarding the question whether this particular attack is 
covered in the original security proofs. Only that much: of course 
the SWAP operation is a valid interaction of Eve with the signal, 
and thus should be covered by the original security proofs, as long 
as those security proofs also allowed a coherent quantum mechanical 
interaction between ancilla and the signal on both directions, just 
as the authors of the comment pointed out. The paragraph in the reply 
to the comment (page 3) \cite{pavicic-nrl-17-response} is a mystery
to me.
 
What happens if Eve attacks only a fraction of the signals with that
particular attack? Yes, no error correction needed ($I_{AE} =1$) and
from the disturbance $D$ in the CM mode one can prove that Eve does not 
learn more than a fraction $f$ of the messages, so $I_{AE}=f$ . As a 
result, the key rate would be $r=1-f$. Of course we don't need to 
know which bits Eve knows and which she does not know. Privacy 
amplification can take care of it, similar as how it takes care of 
signals that Eve could have learned from multi-photon signals in 
weak coherent pulse QKD. There she also learns a fraction of signals,
and Alice and Bob don't know where those bits are located in the 
raw key, and still can distill a secret key.

Note that in his response to the comment the author alludes to 
'the exponential loss in fiber', apparently thinking that this loss, 
as with multi-photon signals, might eventually allow Eve to make 
sure that only signals where she knows the outcome remain in the raw 
data. However, that does not work in the particular case. Eve could 
not prevent Alice and Bob form performing the CM part of the protocol,
hiding that part in the loss. Eve cannot distinguish between CM and 
MM signals, and thus cannot selectively suppress one or the other 
(quite in contrast to multi-photon signals in WCP BB84). Note that no
corresponding argument is made in the original article, though there 
are remarks made, such as on page 3, right column:

¿Here, it should be stressed that photons in LM05 cover twice the 
distance they cover in BB84. So, if the probability of a photon to 
be detected over only Bob-Alice distance is p, the probability of 
being detected over Bob-Alice-Bob distance will be p2 and Eve would 
be able to hide herself in CM exponentially better than in BB84.¿

I cannot follow that logic. As said before, there is no handle for 
Eve to suppress signals selectively in CM mode. The author does not 
give a clear way how this attack angle could be exploited. (And I 
am sure there isn't any way to exploit it.)

I am afraid that the author misunderstood the security framework 
that stands behind the formula $r=I_{AB}-I_{AE}$ in a fundamental way, 
and thus that his conclusions are deeply flawed and invalid in a 
very apparent way.

\section{Author's Response to the Report}
\label{response}

The whole report is busy with $r=I_{AB}-I_{AE}$. It claims (2nd
paragraph from the bottom) - when Eve attacks only a fraction of
signals - that (a) $I_{AB}=1$ (she/he writes $I_{AE}$, but this is a
misprint) and (b) ``from the disturbance $D$ in the CM (control mode)
Eve does not learn more than a fraction $f$ of the messages, so
$I_{AE}=f$. As a result, the key rate would be $r=1-f$." Then it
claims that I did not take all that into account and that I
misunderstood $r=I_{AB}-I_{AE}$.

But in Fig. 5(b), on p. 4 of \cite{pavicic-nrl-17}, I plotted
$I_{AE}(D)$, stressing that ``$D$ is the disturbance in the CM
(control mode).'' Also, in Conclusions, p. 5, last paragraph of
\cite{pavicic-nrl-17} ``Eve induces a disturbance ($D$) only in
the control mode (CM) and therefore the standard approach and
protocols for estimating and calculating the security are not
available since they all assume the presence of $D$ in MM
(message mode)."

Yet, the referee claims immediately afterwards that ``we don't need
to know which bits Eve knows and which she does not know. Privacy
amplification (PA) can take care of it.''

But PA for the considered attack has never been explicitly
calculated. And on p. 5, left column, bottom and right column, top
of \cite{pavicic-nrl-17} I write: "The only procedure we are left
with to establish the security is the privacy amplification. When
Eve possesses just a fraction of data, she will loose trace of her
bits and Alice and Bob's ones will shrink. Eve might be able to
recover data by guessing the bits she misses and reintroduces all
bits again in the hash function. If unsuccessful, her information
will be partly wiped away. However, Alice and Bob meet a crucial
problem with designing their security procedure (e.g., hash function)
which would guarantee that Eve is left with no information about the
final key. They do not have a critical amount of Eve's bits"
which would tell them at which $D$ in the CM they have to abort the
transmission, and this is the main conclusion of my paper.

So, the ``explanation'' of the forceful retraction is nonsensical:
``peer review has revealed a flaw in the application of the key
rate equation $r=I_{AB}-I_{AE}$. For calculation of the term $I_{AE}$,
the effect of disturbance ($D$) on both the message mode (MM) and
control mode (CM) was not taken into account.''

\section{Discussion}
\label{disc}

We all agree that

\begin{enumerate}[(i)]
\item There is no disturbance ($D$) in the message mode (MM), only
  in the control mode (CM);
\item Under the man-in-the-middle attack we always have $I_{AB}=1$
  and therefore LM05 protocol, ``in effect, [amounts to] sending
  plain text via MM secured by occasional verification of photon
  states in CM''
  \cite[p.~5, right column, 2nd paragraph]{pavicic-nrl-17};
\item The only available procedure to sift out Eve's bits is the
  privacy amplification.
\end{enumerate}

There are two disagreements, though.

\begin{enumerate}
\item
  The editors, the anonymised referee, and the authors of
\cite{shaari-mancini-pirandola-lucamarini-18-v1,shaari-mancini-pirandola-lucamarini-18-v2,shaari-mancini-pirandola-lucamarini-18-v3}
claim that something is wrong and deeply flawed with the way I handle
$r=I_{AB}-I_{AE}$ but they do not say what is wrong. They only say
(see Sec.~\ref{intro}) that ``the effect of disturbance ($D$) on
both the message mode (MM) and control mode (CM) was not taken into
account.'' But there is no $D$ in MM, as we all agree, and in
Fig.~5(b) of \cite{pavicic-nrl-17} I plot $I_{AE}(D)$, i.e., $I_{AE}$
as a function of $D$ in CM. Hence, the accusation is void
and unscientific. 
\item
The anonymised referee writes in his report above: ``from the
disturbance $D$ in the CM mode one can prove that Eve does not 
learn more than a fraction $f$ of the messages, so $I_{AE}=f$ .
As a result, the key rate would be $r=1-f$. Of course we don't
need to know which bits Eve knows and which she does not know.
Privacy amplification can take care of it, similar as how it takes
care of signals that Eve could have learned from multi-photon signals
in weak coherent pulse QKD. There she also learns a fraction of
signals, and Alice and Bob don't know where those bits are located
in the raw key, and still can distill a secret key.''\\
The problem with this is that the referee does not tell us {\em how}
``privacy amplification can take care of it.'' When Eve is in
the line all the time ($D=0.5$), then privacy amplification obviously 
does not work since then Bob-Alice's key and Eve's key are
identical and Eve's bits cannot be distilled out. So, is it
feasible for $D=0.499$? Or for $D=0.3$? Or for $D=0.2$?
Or for $D=0.11$? Has anyone carried out such
an analysis? No one to my knowledge. 
\end{enumerate}

And that is why the main point of my paper is, as
I stressed in my response above, that we do not have a procedure
``which would tell Alice [and Bob] at which $D$ in the CM they
have to abort the transmission, and this is the main conclusion of
my paper.'' 

Consequently, I cannot agree to this unfounded, unscientific,
and rather Kafkian \cite{kafka-trial} retraction.

I put together all this as a service to community and readers.

\end{document}